\documentclass[aps,prmater,reprint,superscriptaddress]{revtex4-1}

\usepackage{graphicx}% Include figure files
\usepackage{dcolumn}% Align table columns on decimal point
\usepackage{bm}% bold mathManuscript Title:\\

\usepackage{upgreek}
\usepackage{amsmath}

\usepackage[utf8]{inputenc}
\usepackage[english]{babel}
\usepackage{color}

\begin{document}

\title{Pattern formation during deformation of metallic nanolaminates}

\author{Adrien Gola}
\affiliation{Department of Microsystems Engineering, University of Freiburg, Georges-K\"ohler-Allee 103, 79110 Freiburg, Germany}
\affiliation{Institute for Applied Materials, Karlsruhe Institute of Technology, Kaiserstra\ss e 12, 76131 Karlsruhe, Germany}

\author{Ruth Schwaiger}
\affiliation{Institute for Applied Materials, Karlsruhe Institute of Technology, Kaiserstra\ss e 12, 76131 Karlsruhe, Germany}

\author{Peter Gumbsch}
\affiliation{Institute for Applied Materials, Karlsruhe Institute of Technology, Kaiserstra\ss e 12, 76131 Karlsruhe, Germany}
\affiliation{Fraunhofer IWM, W\"ohlerstra\ss e 11, 79108 Freiburg, Germany}

\author{Lars Pastewka}
\email[]{lars.pastewka@imtek.uni-freiburg.de}
\affiliation{Department of Microsystems Engineering, University of Freiburg, Georges-K\"ohler-Allee 103, 79110 Freiburg, Germany}
\affiliation{Freiburg Materials Research Center, University of Freiburg, Stefan-Meier-Stra\ss e 21, 79104 Freiburg, Germany}
\affiliation{Cluster of Excellence livMatS @ FIT --- Freiburg Center for Interactive Materials and Bioinspired Technologies, University of Freiburg, Georges-K\"ohler-Allee 105, 79110 Freiburg, Germany}

\date{\today}

\begin{abstract}
We used nonequilibrium molecular dynamics simulations to study the shear deformation of metallic composites composed of alternating layers of Cu and Au. Our simulations reveal the formation of ``vortices'' or ``swirls'' if the bimaterial interfaces are atomically rough and if none of the $\{111\}$ planes that accommodate slip in fcc materials is exactly parallel to this interface. We trace the formation of these patterns back to grain rotation, induced by hindering dislocations from crossing the bimaterial interface. The instability is accompanied by shear-softening of the material. These calculations shed new light on recent observations of pattern formation in plastic flow, mechanical mixing of materials and the common formation of a tribomutation layer in tribologically loaded systems.
\end{abstract}

%\keywords{}

\maketitle

\section{Introduction}

Nanolaminates or multilayers are layered composites, consisting of planar layers of alternating composition with thickness on the order of nanometers to hundreds of nanometers. They have attracted interest over the past decades for their excellent mechanical properties such as high strength and wear resistance~\cite{Clemens:1999:20,Misra:2001:217,Freund:2004:}.
In addition, nanolaminates are interesting model systems because the full deformation field can be extracted in experiments simply by tracing the layer structure \emph{post-mortem}. 

Significant plastic deformation takes place when indenting or scratching a surface~\cite{gola_scratching_2019}. Indeed, plastic deformation is responsible for part of the material loss during abrasive~\cite{Khruschov:1974:69} and sliding wear~\cite{Rigney:1979:345}. The plowing motion of asperities on the counter body contributes to the friction between two materials~\cite{Rigney:1979:345, Mishra:2012:417, Mishra:2012:45452}. Nanolaminates allow the tracing of the subsurface deformation through the lifetime of a frictional contact, which may reveal the contribution of plasticity to these processes.

\begin{figure*}
    \includegraphics[width=\textwidth,keepaspectratio]{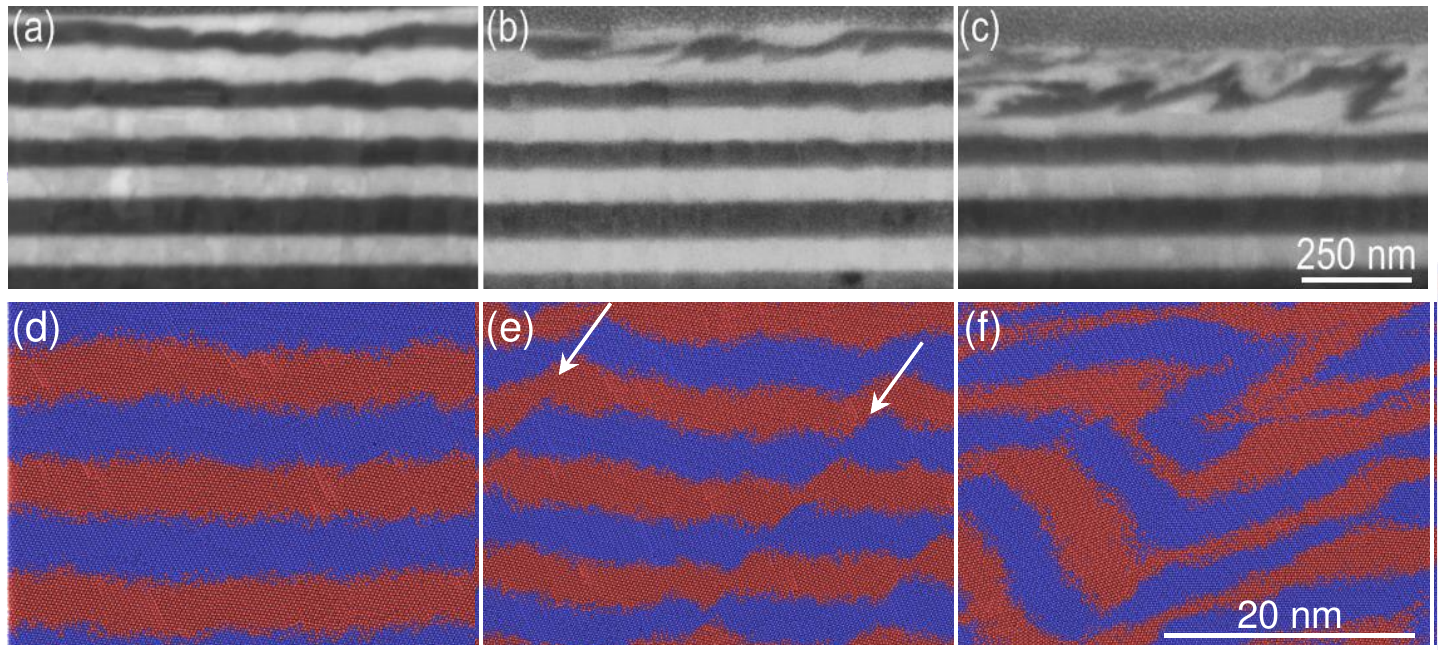}
    \caption{\label{fig:cuau_deformation_atoms_0_5} (a-c) SEM images (at $52^\circ$ sample tilt) of cross sections of Cu$\vert$Au nanolaminates of $100$~nm layer thickness during cyclic sliding with a spherical diamond tip of $16.7~\mu$m diameter. Snapshots are obtained after (a) $10$, (b) $20$ and (c) $100$ sliding cycles. For more details on the experimental procedure see \citet{Luo:2015:67}. (d-f) Snapshots of the molecular dynamics simulation of the nanolaminate with surface planes misoriented by $\theta=5^{\circ}$ with respect to the $(111)$ direction. Snapshots are shown (d) during equi-biaxial deformation along the x and y axis at an applied strain of $\varepsilon_{xx},\varepsilon_{yy}=0.12$, (e) after equi-biaxial deformation to an applied strain $\varepsilon_{xx},\varepsilon_{yy}=0.25$  and (f) after subsequent simple shear deformation to an applied strain $\varepsilon_{xz}=3.0$.}
\end{figure*}

Recently, \citet{Luo:2015:67} have experimentally observed intermixing of individual layers during reciprocating sliding. Their experiments on Cu$\vert$Au nanolaminates with 100~nm thick layers show interface roughening during the early stages of cycling, followed by intermixing and vortex formation (see Fig.~\ref{fig:cuau_deformation_atoms_0_5}a-c). Early molecular dynamics (MD) simulations also showed the formation of vortices of a few atomic distances in size at the sliding contact between metallic crystalline~\cite{Kim:2009:1130,Rigney:2010:3} or amorphous~\cite{fu_sliding_2001,fu_sliding_2003,wu_tribological_2005} bodies.
The vortices in these calculations drove chemical mixing~\cite{odunuga_forced_2005,ashkenazy_shear_2012,zhou_stability_2014,li_glass_2019} and were a consequence of shear localization. 
Direct experimental observation of such vortices had proven difficult but is now possible in metallic nanolaminates where tracing the deformation field is possible at the bimaterial interfaces (Fig.~\ref{fig:cuau_deformation_atoms_0_5}a-c). 
The ``vortices'' or ``swirls'' occurring at the interface are reminiscent of similar patterns observed, for example, during the formation of clouds, from fluid instabilities such as the one named after Lord Kelvin and Hermann von Helmholtz~\cite{hermann_von_helmholtz_uber_1868,thompson_hydrokinetic_1871}. These Kelvin-Helmholtz instabilities occur in the turbulent regime of fluids, i.e. at large Reynolds numbers~\cite{Thorpe:1971:299}. 

More recently, \citet{Pouryazdan:2017:1611} observed similar vortices in high pressure torsion experiments of alternatively stacked Cu$\vert$Al foils of 25~$\mu$m thickness, which were much thicker than the Cu$\vert$Au nanolaminates of \citet{Luo:2015:67}.
\citet{Pouryazdan:2017:1611} pointed out that the velocities required to be in a turbulent regime at the contact of two metallic crystals were unrealistic (hundreds of km s$^{-1}$). 
The Kelvin-Helmholtz instabilities are therefore unlikely to occur in such composites.
\citet{Pouryazdan:2017:1611} proposed an alternative fluid-mechanical model, a layered composite with layers of different viscosity, to demonstrate the formation of vortices at the interface of the two viscous fluids.
While this model did not involve unrealistic flow velocities, it required strains up to 400 for the vortices to fully develop.
Such extremely high applied strains are easily achievable by high-pressure torsion as the shear strain scales linearly with the sample diameter and the number of revolutions.
However, these strains are much larger than those achieved under a sliding track~\cite{Luo:2015:67}.

The mechanism responsible for the formation of patterns such as those shown in Fig.~\ref{fig:cuau_deformation_atoms_0_5} remains unclear.
Mechanical mixing leads to vortices on the atomic scale, much smaller than the patterns shown in Fig.~\ref{fig:cuau_deformation_atoms_0_5} that are clearly on an intermediate scale -- that of the thickness of the individual layers. A purely fluid-mechanical explanation requires unrealistic velocities or extraordinarily large strains. An alternative mechanism for the formation of vortex structures in nanolaminates must therefore be active. 
We investigated the phenomenon of vortex formation using nonequlibrium MD calculations of deformation of Cu$\vert$Au nanolaminates.

Understanding the deformation process can yield insights into ``tribomutation''~\cite{Rigney:2010:3} processes that occur in almost all tribological contacts, where the near-surface material transforms - e.g. by grain refinement,~\cite{emge_effect_2007,stoyanov_experimental_2013} or amorphization~\cite{minowa_stress-induced_1992,pastewka_anisotropic_2011,moras_shear_2018}.

% ----------------
% ---- Methods ---
% ----------------
\section{Simulation model \& methods}

\begin{figure}
    \includegraphics[width=\columnwidth,keepaspectratio]{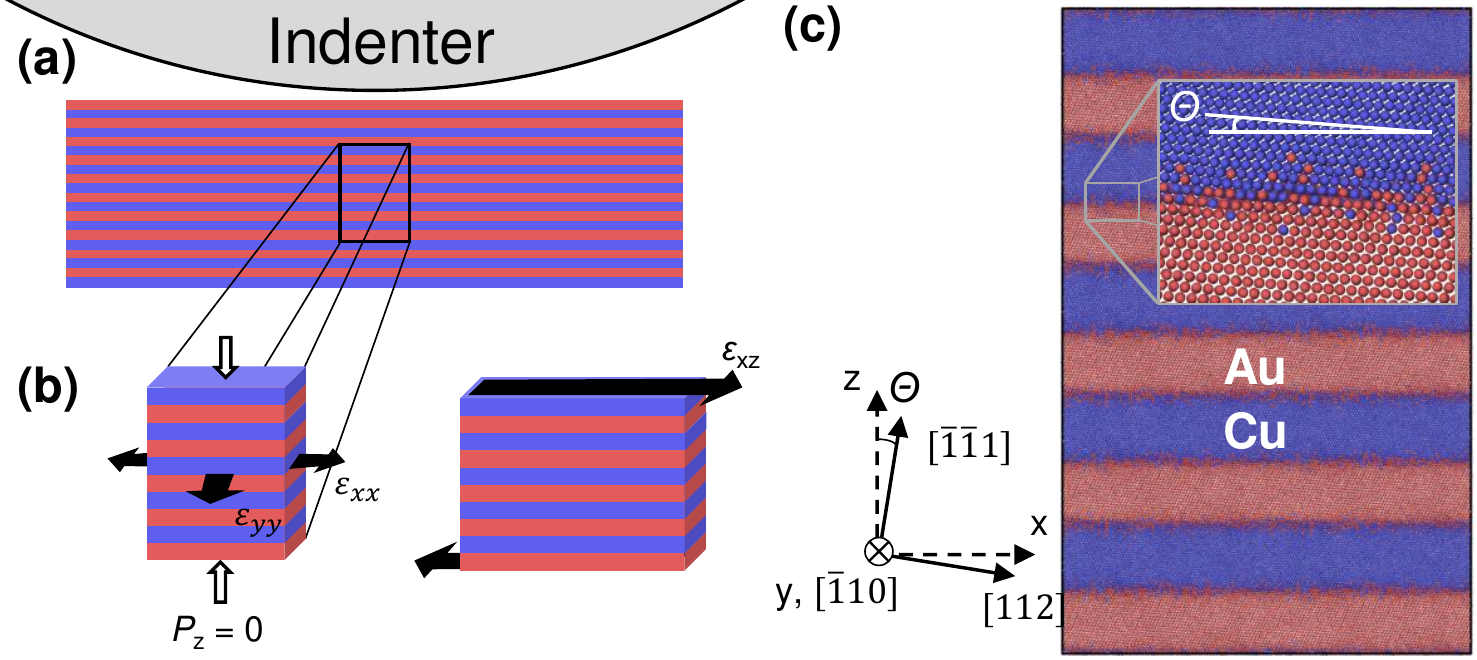}
    \caption{\label{fig:setup} (a) Schematic representation of the experimental setup similar to the one described in~\citet{Luo:2015:67}. (b) Schematic representation of the two deformation steps applied to the representative volume element used in our simulations. (c) Snapshot of the initial simulation setup. The inset shows the misalignment of the lattice. The angle relative to a perfect $(111)$ orientation is denoted by $\theta$.}
\end{figure}

Our simulations are designed to uncover the conditions required for inducing vortex instabilities. Since we know the conditions under which they are observed experimentally, our simulation model is guided by the experiments of \citet{Luo:2015:67} on Cu$\vert$Au nanolaminates. The experiments used a spherical indenter of radius $R$ much larger than the layer thickness $\lambda$, $\lambda/R = 6\cdot 10^{-3}$ in Ref.~\onlinecite{Luo:2015:67} . The stress field is roughly homogeneous on the scale of the layer thickness $\lambda$ and we will only regard representative volume elements subjected to strain-controlled deformation. (See Fig.~\ref{fig:setup}; a structural mechanical model on the full scale of the experiment is out of reach on present-day computers within molecular simulations.) The representative volume element used in this work is shown in Fig.~\ref{fig:setup}c. Due to the film growth process, the surface normal of the bimaterial interface is always close to a $\{111\}$ plane. Our calculations also allow the possibility of having a slight misorientation (angle $\theta$) between the crystallographic direction and the bimaterial interface (see inset of Fig.~\ref{fig:setup}c).

The initial system shown in Fig.~\ref{fig:setup}c is composed of 10 layers of 5~nm thickness each, with an in-plane size of approximately $30\times30$~nm$^{2}$. 
We also investigated possible size effects by running a similar calculation on a ``supercell'' system nine times larger comprised of 30 layers obtained by replicating the undeformed system cell in the $x$- and $z$-directions for $\theta=5^{\circ}$.
Table~\ref{tab:cuau_instabilities_setup} gives the exact number of Au and Cu unit cells used in the in-plane directions and the corresponding tilt angles. We used periodic boundary conditions in all directions.

\begin{table}%[H] add [H] placement to break table across pages
\caption{\label{tab:cuau_instabilities_setup}Minimum simulation cell size used for our simulations. The numbers $n$ and $m$ denote the number of unit cells of the Au and Cu layer, respectively.	A ratio $m/n \not= 1$ and different tilt angles $\theta$ are necessary to accommodate the nominal lattice mismatch and comply with the periodic boundary conditions. All systems are composed of 10 layers, with the exception of the system denoted by supercell that contained 30 layers.}\begin{ruledtabular}
\begin{tabular}{llll}
Setup name & $\theta_\text{Au}/\theta_\text{Cu}\,(^{\circ})$ & $n_{\left[112\right]}/m_{\left[112\right]}$ & $n_{\left[\bar{1}10\right]}/m_{\left[\bar{1}10\right]}$ \\
		\hline
		$0^{\circ}$	  & $0/0$       & $66/74$ &  $125/140$ \\
		$2.5^{\circ}$ & $2.45/2.19$ & $66/74$& $125/140$  \\
		$5^{\circ}$   & $4.90/4.37$ &  $66/74$&  $125/140$ \\
		$5^{\circ}$ (supercell)   & $4.90/4.37$ &  $198/222$&  $125/140$ \\
		$10^{\circ}$   & $10.0/8.93$ & $64/72$ & $125/140$  \\
\end{tabular}
\end{ruledtabular}
\end{table}

We created the initial configuration from ideal Cu and Au fcc slabs with intermixed interfaces. The interfaces were intermixed by randomly flipping Cu and Au atoms over a finite interface width of $15$\AA, such that the final concentration profile follows the error function predicted by simple Fickian interdiffusion~\cite{Gola:845721}. The systems were annealed at $1000$~K for a total of $2$~ns in MD and then quenched down to $300$K at a rate of $350\,\textrm{K ns}^{-1}$. 
This was followed by further aging the system at $300$~K for $1$~ns. 
During these annealing simulations, the temperature was controlled using a Langevin thermostat with a relaxation time of $1$~ps. We used an anisotropic Berendsen barostat~\citep{Berendsen:1984:3684} with a relaxation time of $\sim 5$~ps to maintain zero stress along the $\left[112\right]$ and $\left[110\right]$ directions. All calculation steps were performed with a timestep of $5$~fs.

We replicated the system along the $z$-axis to obtain the final 10 layers as shown in Fig.~\ref{fig:setup}.
The systems then underwent two successive deformation steps: First, equi-biaxial tensile deformation up to a biaxial strain of 25\% followed by simple shear (see Fig.~\ref{fig:setup}b).
All deformation simulations were carried out at 300~K with a strain rate of $10^{8}\:\textrm{s}^{-1}$.
The temperature was controlled using a Nos\'e-Hoover thermostat with a relaxation time of $1$~ps. For simple shear deformation the thermostat is applied only in the direction perpendicular to the shearing direction. 
The equi-biaxial tensile deformation was imposed by changing the simulation box size along the $x$ and $y$ directions according to the defined strain rate. The pressure perpendicular to the deformation (i.e. $z$ direction) was kept at zero using the Andersen/Parinello-Rahman barostat~\cite{andersen_molecular_1980,Parrinello:1981:7182} with a relaxation time constant of 10~ps. 
Simple shear deformation was then applied along the interface planes, i.e. along $[112](\bar{1}\bar{1}1)$ for $\theta = 0^{\circ}$, by homogeneously deforming the box at the prescribed strain rate. Our notation $[abc](hkl)$ for simple shear reports both the direction of shear $[abc]$ and the plane of shear $(hkl)$.

During shear, we quantify the lattice orientation of local regions within our simulation. The lattice orientation was obtained using polyhedral template matching~\cite{Larsen:2016:55007,Stukowski:2010:15012} (PTM). We determined the orientation along the $z$-axis (multilayer growth axis) and used an equal-area projection to assign a color to each atom.  
We also quantify elastic and plastic rearrangements rather than defects by computing the local strain tensor from the analysis of \citet{Falk:1998:7192} within local neighbor spheres of a radius that include just an atom’s nearest neighbors. 
This technique computes the atomic neighborhood in a reference frame and then extracts the deformation gradient tensor $\mathbf{F}_i$ necessary to transform the vectors connecting the atom $i$ of interest to its reference neighborhood to the deformed configuration in a least-squares sense. 
Then from the deformation gradient the local Green-Lagrangian strain tensor is computed~\citep{Shimizu:2007:2923}, $\boldsymbol{\gamma}_i = (\mathbf{F}_i^{T}\mathbf{F}_i - \mathbf{I})/2$.
With this method one can visualize the total amount of local deformation a system has experienced, i.e. where dislocations have passed.

% -------------------------------
% ---- Results and discussion ---
% -------------------------------
\section{Results}

\begin{figure}
    \includegraphics[width=\columnwidth,keepaspectratio]{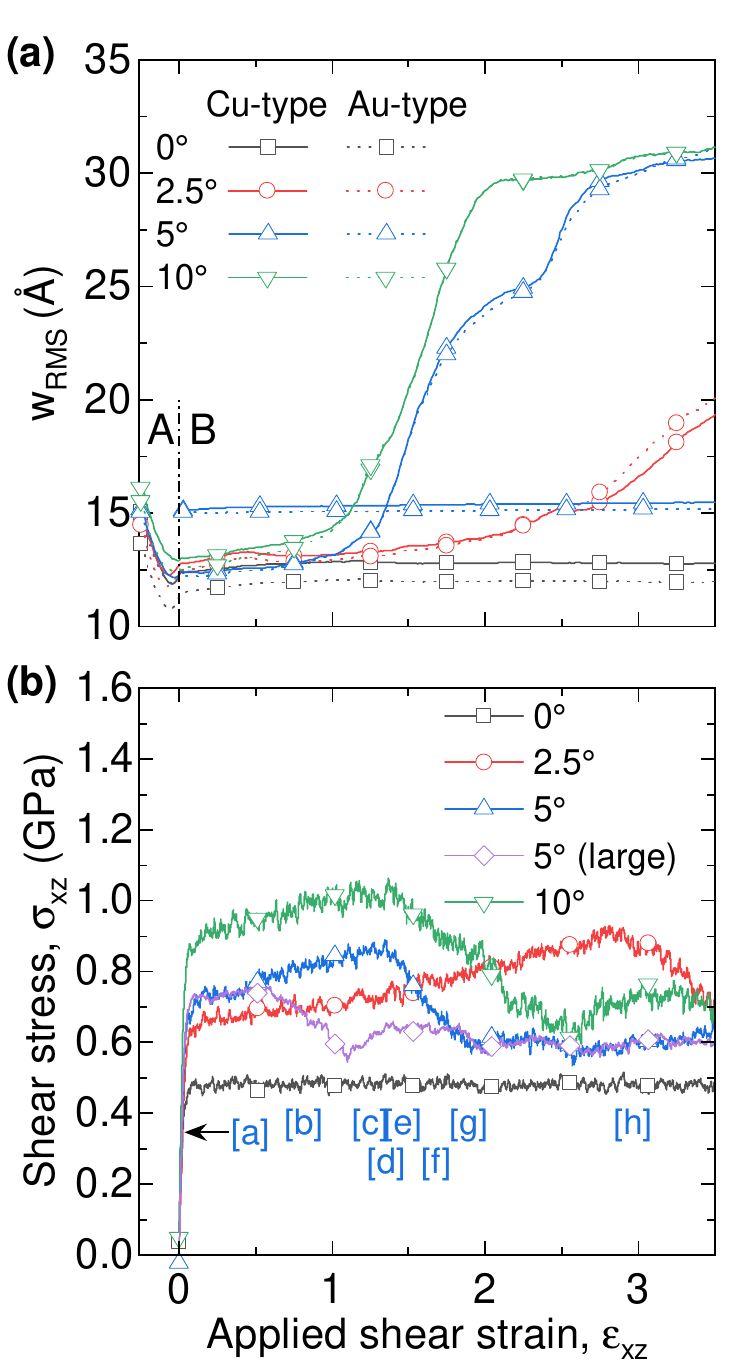}
    \caption{\label{fig:cuau_deformation_wrms_stress}(a) Averaged root mean square of the layer width, $w_{\textrm{RMS}}^{l}$ (Eq.~\eqref{eq:wrms}), over the course of the biaxial (subplot A) and simple shear deformation (subplot B). (b) Stress-strain curve obtained during the simple shear calculation of the misoriented $\left\{111\right\}$ planes for $\theta=0^{\circ},2.5^{\circ},5^{\circ},10^{\circ}$. The blue letters in square brackets indicate the strain values corresponding to the snapshots shown in Fig.~\ref{fig:cuau_shear_5_misorientation} for the $\theta=5^{\circ}$ calculation.} 
\end{figure}

\emph{Equi-biaxial deformation.}
The initial systems with $10$ layers and misalignment ranging from $\theta=0^{\circ}$ to $10^{\circ}$ with interfaces parallel to the $xy$-plane were deformed by equi-biaxial deformation along the $x$ and $y$ axis (see Fig.~\ref{fig:setup}b). 
Each atom was assigned to a single layer $l$ at the beginning of the calculation and this assignment stayed fixed during the course of the calculation.
We characterize the thickness of each layer by its root mean square width,
\begin{equation}
           w_{\textrm{RMS}}^{l}(t) = \sqrt{\frac{1}{N_{l}} \sum_{i \in l} \left( z_{i}(t) - \frac{1}{N_{l}} \sum_{j \in l} z_{j}(t) \right)^{2}},
           \label{eq:wrms}
\end{equation}
with the total number of atoms in this layer, $N_{l}$, and the z-component of atom $i$ at time step $t$, $z_{i}(t)$.
Figure~\ref{fig:cuau_deformation_wrms_stress}a shows the evolution of the average $w_{\textrm{RMS}}^{l}$, averaged over Cu and Au layers $l$ separately.

This measure allows us to track layer thinning and broadening.
During the equi-biaxial deformation, $w_{\textrm{RMS}}$ showed a similar behavior for all the systems independent of their misorientation. 
$w_{\textrm{RMS}}$ monotonously decreased until it reached a minimum at a strain of approximately 20\% before slightly increasing towards the final applied strain of $25\%$. Both Cu-type and Au-type layers thinned down at the same rate.

\begin{figure*}
    \includegraphics[width=\textwidth,keepaspectratio]{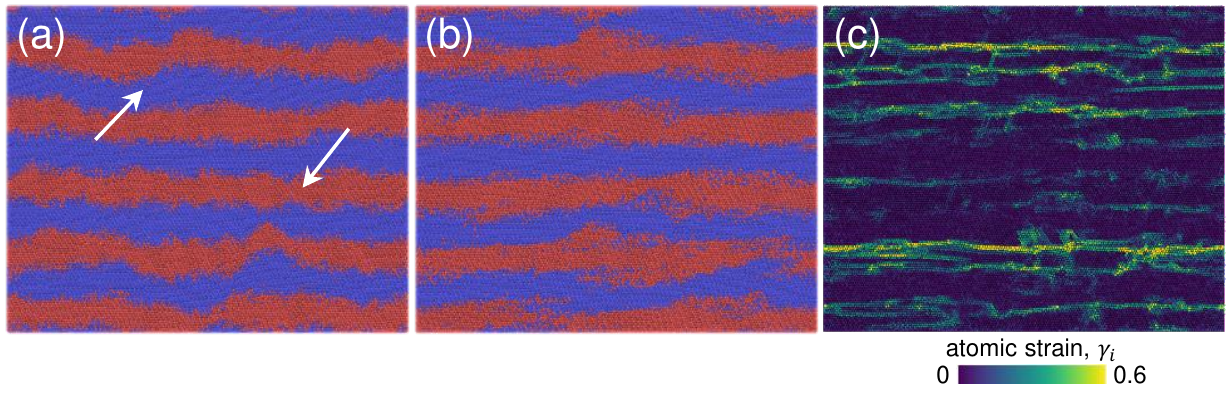}
    \caption{\label{fig:no_misalignment}Snapshots of the nanolaminate stack with $[\bar{1}\bar{1}1]$ planes initially aligned with the interfaces. (a) snapshot taken after equi-biaxial deformation along the x and y axis to $\varepsilon_{xx}=\varepsilon_{yy}=0.25$, and (b,c) after subsequent simple shear deformation to $\varepsilon_{xz}=3.0$.
    Atoms in panel (a) and (b) are colored according to their type, Cu atoms are in blue and Au atoms in red. Atoms in panel (c) are colored after their local atomic shear strain $\gamma_{i}$. The atomic strains are computed based on reference system at an applied strain $\varepsilon_{xz}=2.9$.} 
\end{figure*}

Figures~\ref{fig:cuau_deformation_atoms_0_5}e (lattice tilt of $\theta=5^{\circ}$) and~\ref{fig:no_misalignment}a ($\theta=0^{\circ}$) show snapshots of the systems after equi-biaxial deformation at 25\% strain. From these images it is clear that the initial straight interfaces roughened during deformation. Both $\theta=5^{\circ}$ and $0^{\circ}$ systems developed shear bands that crossed several layers, creating the most pronounced roughness features at the interfaces (shown by white arrows in the corresponding figures).

\emph{Simple shear deformation.}
After equi-biaxial deformation, the systems were deformed by applying simple shear. Figure~\ref{fig:cuau_deformation_wrms_stress}b shows the stress strain curves obtained for the four systems of interest. 
The system without misorientation ($\theta = 0^{\circ}$) showed a flat stress strain curve. This behavior is identical to the one observed for simple shear deformation of a perfectly flat Cu$\vert$Au bilayer sheared parallel to the interface (see supporting material of Ref.~\onlinecite{Gola:2019:}). After the initial elastic response, both systems yielded at around 0.5~GPa and then entered a flow regime with a flow stress fluctuating around this value. For flat interfaces, an analysis of the local strain rate~\cite{falk_dynamics_1998} revealed that all of the strain is accommodated at the bilayer interfaces.

The stress required to deform the systems with misorientation of the $\left\{111\right\}$ planes increased with increasing $\theta$. The nanolaminates yielded between 0.6~GPa for $\theta=2.5^{\circ}$ and 0.88~GPa for $\theta=10^{\circ}$. All misoriented systems then showed strain hardening, manifested in an almost linear stress-strain relationship. During hardening, the systems encountered an instability. The shear stress started to drop at $\sigma\approx0.9$~GPa (at an applied strain of $\varepsilon\approx2.9$) for $\theta=2.5^{\circ}$, $\sigma\approx 0.9$~GPa ($\varepsilon\approx1.3$) for $\theta=5^{\circ}$, and $\sigma\approx1.0$~GPa ($\varepsilon\approx1.3$) for $\theta=10^{\circ}$. 
The shear stress eventually dropped to roughly the same value of 0.6~GPa for the systems. The systems continued to shear at that stress during subsequent deformation.
Again, analysis of the local strain rate showed that strain was accommodated at localized regions oriented parallel to the simulation cell boundary once the instability had occurred.

Figure~\ref{fig:cuau_deformation_wrms_stress}a shows that the point of the instability (the stress drop) for the misoriented systems coincided with the rapid increase of  $w_{\textrm{RMS}}$, while for $\theta=0^{\circ}$ (no instability), $w_{\textrm{RMS}}$ stayed constant over the whole shear deformation. 
Figure~\ref{fig:cuau_deformation_atoms_0_5}f shows a snapshot of the system with $\theta=5^{\circ}$ sheared to an applied strain of $\varepsilon=3.0$. For $\theta=5^{\circ}$, the layered structure is severely deformed with noticeable patterns, similar to the experimental ``vortices'' shown in Fig.~\ref{fig:cuau_deformation_atoms_0_5}c. Without misorientation ($\theta=0^{\circ}$) the deformation is accommodated along the interfaces and the bimaterial interfaces remain straight (see Fig.~\ref{fig:no_misalignment}b and c).

\begin{figure*}
    \includegraphics[width=\textwidth,keepaspectratio]{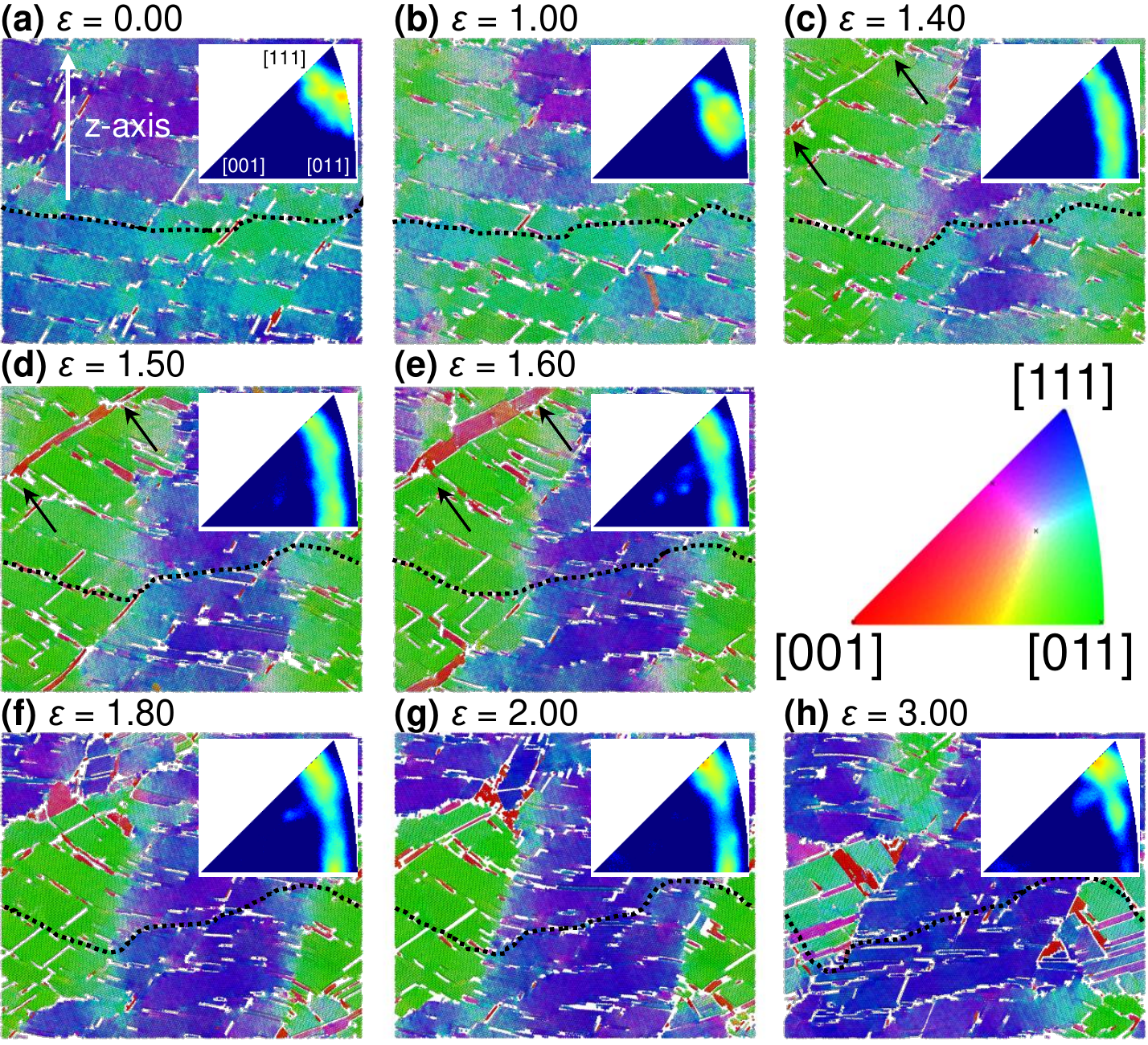}
    \caption{\label{fig:cuau_shear_5_misorientation}Orientation maps of the system with a misorientation of $\theta=5^{\circ}$ during simple shear deformation. Non-FCC atoms have been removed, FCC atoms are color coded after their local lattice orientations along the z-axis. The interfaces are less visible with this color scheme, for clarity one interface is marked up with a black dashed line. The arrows in (c) indicate the boundaries of the twinning event.
    The insets show inverse pole figures with the local lattice orientation density along the z-axis of the simulation cell in the standard stereographic triangle. The color coding used here follow a jet colormap with high pole density in red and low density in blue.
    }
\end{figure*}

In order to gain more insights into the mechanism underlying the observed instability and the formation of the vortex-like patterns, we now focus on the system with a misorientation of $\theta=5^{\circ}$ and describe in detail the deformation process. Figure~\ref{fig:cuau_shear_5_misorientation} shows snapshots of the system. The atoms are color-coded according to their local lattice orientation in the direction of the $z$-axis (normal to the initial bilayer interfaces). Atoms that are not in a local fcc environment (and can hence not be assigned an orientation) are not shown. The insets in Fig.~\ref{fig:cuau_shear_5_misorientation} show inverse pole figures of the distribution of these local lattice orientations found throughout the simulation cell.

Before shear there were two main orientations in the system (Fig.~\ref{fig:cuau_shear_5_misorientation}a), colored by purple and green, corresponding to orientations of the $z$-axis approximately along the $[432]$ and $[553]$ directions, respectively. Note that the transition between these two regions roughly corresponds to the location of the shear bands seen in Fig.~\ref{fig:cuau_deformation_atoms_0_5}e.
At a shear strain of $\varepsilon = 1.4$, i.e. right after the stress drops, we can see traces of dislocations that cross three layers, marked by the two black arrows in Fig.~\ref{fig:cuau_shear_5_misorientation}c. 
At this point we also observed a significant rotation of the local lattice toward the $[011]$ direction, leading to a broad band of orientations in the inverse pole figure (inset to Fig.~\ref{fig:cuau_shear_5_misorientation}c). 
At $\varepsilon = 1.6$, this interlayer region had widened (marked with two black arrows). We now had in the system the presence of a large twinned area (red in Fig.~\ref{fig:cuau_shear_5_misorientation}c-e, marked by two black arrows). 

At an applied strain of $\varepsilon = 1.8$ and above, this zone shrunk as part of the lattice rotated back to a $(111)$ orientation.  This rotation led to the creation of two misoriented zones in the system, clearly visible by the distinct green and blue zones in Fig.~\ref{fig:cuau_shear_5_misorientation}f-g.

In the final snapshot ($\varepsilon = 3.0$, Fig.~\ref{fig:cuau_shear_5_misorientation}h) there are two distinct orientations in the system, a main one along $[433]$ ($8^{\circ}$ rotation from [111]) and  a smaller grain $[542]$ (rotated), separated by a boundary. 
Further analysis revealed a misorientation of approximately $14^{\circ}$ between the two grains.

The rotation process can also be clearly identified from the insets of Figs.~\ref{fig:cuau_shear_5_misorientation}c-h. After the stress drop, a continuous distribution of orientations between $[111]$ and $[011]$ of the standard stereographic triangle emerged (inset Fig.~\ref{fig:cuau_shear_5_misorientation}c), followed by the appearance of two distinct orientations in the insets of Figs.~\ref{fig:cuau_shear_5_misorientation}c-g. Finally at shear strain of $\varepsilon = 3.0$ (inset Fig.~\ref{fig:cuau_shear_5_misorientation}h), the rotation had completed, i.e. all atoms in the system returned to an orientation close to $[111]$.

\begin{figure}
    \includegraphics[width=\columnwidth,keepaspectratio]{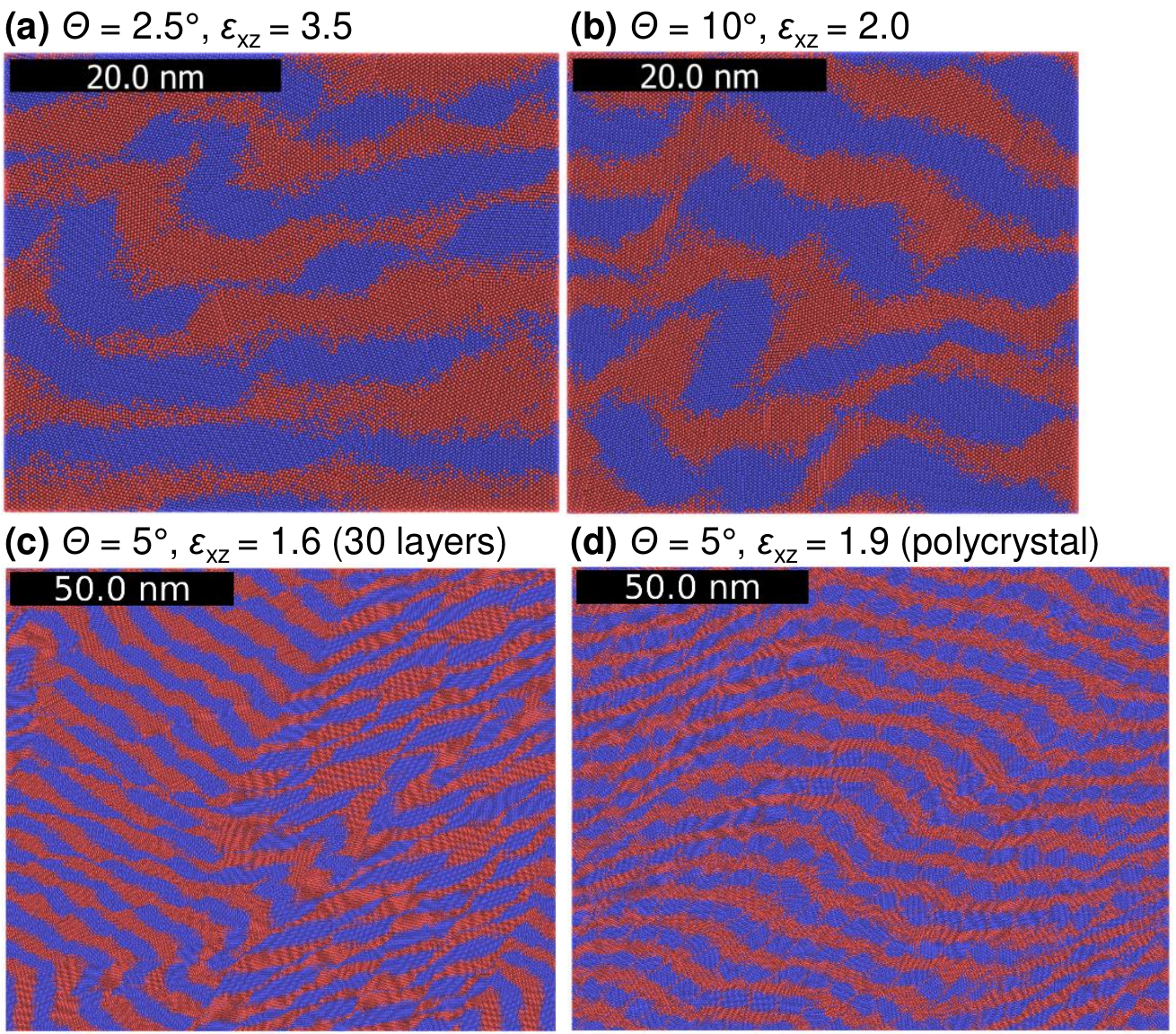}
    \caption{\label{fig:cuau_test_cases}Snapshots of various additional atomistic models. (a) System with $\theta=2.5^{\circ}$ misorientation and $10$ layers at a shear strain of $\varepsilon = 3.5$. (b) System with $\theta=10^{\circ}$ misorientation and $10$ layers at a shear strain of $\varepsilon = 2.0$. (c) System with $\theta=5^{\circ}$ misorientation and 30 layers (a supercell of the smaller system) at a shear strain of $\varepsilon = 1.6$. (d) System with $\theta=5^{\circ}$ misorietation and 30 polycrystalline layers after shear strain of $\varepsilon = 1.9$. Atoms color coded after their type, Cu atoms are in shown blue and Au atoms are in shown in red.}
\end{figure}

We also evaluated the influence of lattice misorientation and system size. Fig.~\ref{fig:cuau_test_cases}a and b show snapshots of the calculations carried out with at a misorientation of $\theta=2.5^\circ$ and $\theta=10^\circ$ deformed using the aforementioned simple shear protocol.
Figure~\ref{fig:cuau_test_cases}c shows results for a larger supercell consisting of $30$ layers, while Fig.~\ref{fig:cuau_test_cases}d shows results for the most realistic of the systems investigated here, a polycrystalline nanolaminate. To set up the polycrystal, we chose hexagonal grains within each layer of lateral size identical to the layer thickness. The grains where randomly rotated around the $z$-axis of our simulation cell (see Fig.~\ref{fig:setup} for the coordinate system).  We observe comparable vortex patterns for all these systems. 
Figure~\ref{fig:cuau_deformation_wrms_stress}b shows the stress-strain curve obtained for the supercell system of Fig.~\ref{fig:cuau_test_cases}c. Compared to the smaller systems, the large system showed the same yield stress and initially the same hardening, but the instability occurred at a smaller strain of $\varepsilon \approx 0.6$, compared to $\varepsilon\approx 1.3$ for the smaller system. Stress was comparable after the instability had occurred.

%At a shear strain of $\varepsilon = 3.0$ the larger system showed similar patterns (Fig.~\ref{fig:cuau_shear_5_size_effect}b). The analysis of the local lattice orientation on Figs.~\ref{fig:cuau_shear_5_size_effect}c-d shows that similarly misoriented areas were present after equi-biaxial deformation along shear bands areas, and after simple shear a clearly rotated region was present.

\section{Discussion}

% discuss biaxial tensile deformation RMS
The initial reduction of the averaged $w_{\textrm{RMS}}^{l}$ in Fig.~\ref{fig:cuau_deformation_wrms_stress}a during the equi-biaxial tensile deformation can easily be connected to the layers co-deforming homogeneously. Even though co-deformation is here dictated by the periodic boundary conditions applied on the system, such behavior is observed experimentally under inhomogeneous loading conditions such as during nanoindentation of Cu$\vert$Au~\cite{Li:2017:20}.
We note the inversion of the averaged $w_{\textrm{RMS}}^{l}$ slope in Fig.~\ref{fig:cuau_deformation_wrms_stress}a that corresponds to the appearance of roughness at the interfaces. Indeed during equi-biaxial tensile deformation, where layers should thin down due to mass conservation, the only way to increase the averaged $w_{\textrm{RMS}}^{l}$ value is for the layer to shear and therefore generate a rough interface. Such shear bands are visible in Fig.~\ref{fig:cuau_deformation_atoms_0_5}. 
This phenomenology of equi-biaxial deformation seems to be independent of misorientation.
Similar shear banding has been observed in different multilayered systems under indentation~\cite{Li:2010:3049,Yan:2013:227}. Roughness develops because materials do not deform continuously like in a laminar flow, but by slip on distinct glide planes~\cite{zaiser_scale_2006}. Similar mechanisms have been reported for creating surface roughness at free surfaces~\cite{hinkle_universal_2019} or buried interfaces~\cite{bellon_nonequilibrium_1995} during deformation.

% discuss relationship rotation and tilting angle
The first effect of misorientation away from the most favourable slip plane $\left\{111\right\}$ is the increase of yield stress with increasing misorientation $\theta$ during the simple shear deformation (parallel to the initial $x$-$y$ interface plane) in Fig.~\ref{fig:cuau_deformation_wrms_stress}b. As observed in previous work under parallel shear to the interface, FCC nanolaminates accommodate the deformation by interface sliding~\cite{GOLA2018236,Gola:2019:}.
 The higher resistance to shear therefore emerges because the interfacial shear strength increases: Any dislocation with a Burgers vector component normal to the interface that sits at the interface provides an obstacle to dislocation gliding parallel and along the interface. In order to accommodate for misorientation, steps must exist at the interfaces. The spacing between the steps is proportional to the misorientation angle $\theta$. Thus, the steps disrupt the flat dislocation network at the interfaces and act as obstacles. The dislocations have to climb to allow for interface sliding. 
\citet{Zhang:2016:194} noted a similar behavior for the Cu$\vert$Nb system under simple shear parallel to the interface. Additionally, the local misorientation re-directs the deformation away from the originally favoured $\left\{111\right\}$ plane. Dislocations on glide planes at an angle to the macroscopic shear direction experience a decrease in the resolved shear stress. This reduction in resolved shear stress is very small in the presently studied interfaces because of the small misorientation. However, together with stronger obstacles, it leads to an increase in the applied stress required for deformation and because of the increase in obstacle density with deformation it also leads to strain hardening.

The second effect of misorientation is visible at larger shear strain. The stress drops for $\theta \neq 0$ in Fig.~\ref{fig:cuau_deformation_wrms_stress}b. This stress drop corresponds to the sudden increase of the averaged $w_{\textrm{RMS}}^{l}$ in Fig.~\ref{fig:cuau_deformation_wrms_stress}a, which can be traced back to the appearance of waviness in the layer shapes as seen in Figs.~\ref{fig:cuau_shear_5_misorientation}a-h.
The stress drops because of a burst of dislocations that cross several layers with the two nucleation points shown by black arrows in Fig.~\ref{fig:cuau_shear_5_misorientation}c. As the strain increases, dislocations continue to glide on successive planes, leading to the growth of a twinned area, in purple in Figs.~\ref{fig:cuau_shear_5_misorientation}c-g. This growth mechanism is compatible with twin growth from a grain boundary with the emission of partial dislocation on successive $\left\{111\right\}$ planes~\cite{hirth1982theory}.
In other words, strain hardening stops once localization of the deformation on an inclined plane releases some of the stored dislocation and stress. These localized slip bands lead to the formation of multiple crystallites that further individually rotate until the next $\left\{111\right\}$ plane is brought parallel to the macroscopic shear direction. The origin of the wave structure observed in the layer structure can be traced down to these localized slip bands.
At larger strain, $\varepsilon > 2$, we attribute the plateau observed for  $w_{\textrm{RMS}}^{l}$ to folding of individual layers as the wavy structure is sheared (shown in Fig.~\ref{fig:cuau_deformation_atoms_0_5}d).

We have tested the relevance of the biaxial deformation to the overall mechanism by shearing directly the $\theta = 5^{\circ}$ system without performing the initial biaxial deformation. Just misaligning the lattice by an angle $\theta$ is not enough to trigger layer rotation and folding. In the contex of the above discussion, this indicates that the shear bands created during biaxial tensile deformation are a necessary disturbance to create distinct obstacles at the interface and points of stress concentration.

The supercell calculations revealed that the instability occurs at smaller strain as the system size increases. This indicates that strain hardening is mostly the result of a size effect acting only on ``small'' systems~\cite{horstemeyer_length_2001}, indicating that the instability needs certain types of defects to nucleate. For a large realization of a nanolaminate, we would therefore expect that nucleation of plastic instabilities that lead to waves, folds and vortices to occur at small strains, likely immediately upon plastic deformation. Note that even the strain required to nucleate these phenomena in our smallest cells are much smaller than the strains reported for a fluid mechanical instability by \citet{Pouryazdan:2017:1611}.

\section{Summary \& Conclusions}

We presented nonequilibrium molecular dynamics calculation of the deformation of Cu$\vert$Au nanolaminates. Our simulations reveal pattern formation during flow, namely the formation of ``waves'', ``vortices'' or ``swirls'' similar to those recently observed in sliding~\cite{Luo:2015:67} and high-pressure torsion~\cite{Pouryazdan:2017:1611} experiments. We identify two crucial ingredients to stabilizing these patterns: First, we need rough bimaterial interfaces, here created by an initial equi-biaxial compression step. Second, the $\{111\}$ planes that accommodate slip in fcc Cu and Au need to be slightly misoriented with respect to the bimaterial interfaces. 
Both factors hinder dislocations from crossing the bimaterial interfaces. This leads to stress concentrations and the formation of crystallites of sizes on the order of the layer thickness that rotate as an alternative mechanism to accommodate strain. This rotation leads to the formation of waves that are folded by subsequent deformation. The instability occurs at small strain and is not of a fluid mechanical nature, such as the often-quoted Kelvin-Helmholtz instability. It is rather intimately tied to the existence of a discrete crystal lattice that deforms by creation, annihilation and motion of crystal defects.

\begin{acknowledgments}
We thank Christian Greiner for helpful comments on the manuscript and Z.P. Luo for the SEM micrographs. The research was partially supported by the Helmholtz Association (HCJRG-217) and the Deutsche Forschungsgemeinschaft DFG (grant PA 2023/2). All our molecular dynamics calculations were carried out with LAMMPS~\cite{Plimpton:1995:1}. ASE~\cite{hjorth_larsen_atomic_2017} and OVITO~\cite{Stukowski:2010:15012} were used for pre-processing, post-processing and visualization. Computations were carried out on NEMO (University of Freiburg, DFG grant INST 39/963-1), ForHLR II (Steinbuch Center for Computing at Karlsruhe Institute of Technology, project ``MULTILAYER'') and JUQUEEN (J\"ulich Supercomputing Center, project ``hka18'').
\end{acknowledgments}

%\bibliography{references}

\begin{thebibliography}{49}%
\makeatletter
\providecommand \@ifxundefined [1]{%
 \@ifx{#1\undefined}
}%
\providecommand \@ifnum [1]{%
 \ifnum #1\expandafter \@firstoftwo
 \else \expandafter \@secondoftwo
 \fi
}%
\providecommand \@ifx [1]{%
 \ifx #1\expandafter \@firstoftwo
 \else \expandafter \@secondoftwo
 \fi
}%
\providecommand \natexlab [1]{#1}%
\providecommand \enquote  [1]{``#1''}%
\providecommand \bibnamefont  [1]{#1}%
\providecommand \bibfnamefont [1]{#1}%
\providecommand \citenamefont [1]{#1}%
\providecommand \href@noop [0]{\@secondoftwo}%
\providecommand \href [0]{\begingroup \@sanitize@url \@href}%
\providecommand \@href[1]{\@@startlink{#1}\@@href}%
\providecommand \@@href[1]{\endgroup#1\@@endlink}%
\providecommand \@sanitize@url [0]{\catcode `\\12\catcode `\$12\catcode
  `\&12\catcode `\#12\catcode `\^12\catcode `\_12\catcode `\%12\relax}%
\providecommand \@@startlink[1]{}%
\providecommand \@@endlink[0]{}%
\providecommand \url  [0]{\begingroup\@sanitize@url \@url }%
\providecommand \@url [1]{\endgroup\@href {#1}{\urlprefix }}%
\providecommand \urlprefix  [0]{URL }%
\providecommand \Eprint [0]{\href }%
\providecommand \doibase [0]{http://dx.doi.org/}%
\providecommand \selectlanguage [0]{\@gobble}%
\providecommand \bibinfo  [0]{\@secondoftwo}%
\providecommand \bibfield  [0]{\@secondoftwo}%
\providecommand \translation [1]{[#1]}%
\providecommand \BibitemOpen [0]{}%
\providecommand \bibitemStop [0]{}%
\providecommand \bibitemNoStop [0]{.\EOS\space}%
\providecommand \EOS [0]{\spacefactor3000\relax}%
\providecommand \BibitemShut  [1]{\csname bibitem#1\endcsname}%
\let\auto@bib@innerbib\@empty
%</preamble>
\bibitem [{\citenamefont {Clemens}\ \emph {et~al.}(1999)\citenamefont
  {Clemens}, \citenamefont {Kung},\ and\ \citenamefont
  {Barnett}}]{Clemens:1999:20}%
  \BibitemOpen
  \bibfield  {author} {\bibinfo {author} {\bibfnamefont {B.~M.}\ \bibnamefont
  {Clemens}}, \bibinfo {author} {\bibfnamefont {H.}~\bibnamefont {Kung}}, \
  and\ \bibinfo {author} {\bibfnamefont {S.~A.}\ \bibnamefont {Barnett}},\
  }\href {\doibase 10.1557/S0883769400051502} {\bibfield  {journal} {\bibinfo
  {journal} {MRS Bull.}\ }\textbf {\bibinfo {volume} {24}},\ \bibinfo {pages}
  {20} (\bibinfo {year} {1999})}\BibitemShut {NoStop}%
\bibitem [{\citenamefont {Misra}\ and\ \citenamefont
  {Krug}(2001)}]{Misra:2001:217}%
  \BibitemOpen
  \bibfield  {author} {\bibinfo {author} {\bibfnamefont {A.}~\bibnamefont
  {Misra}}\ and\ \bibinfo {author} {\bibfnamefont {H.}~\bibnamefont {Krug}},\
  }\href {\doibase 10.1002/1527-2648(200104)3:4<217::AID-ADEM217>3.0.CO;2-5}
  {\bibfield  {journal} {\bibinfo  {journal} {Adv. Eng. Mater.}\ }\textbf
  {\bibinfo {volume} {3}},\ \bibinfo {pages} {217} (\bibinfo {year}
  {2001})}\BibitemShut {NoStop}%
\bibitem [{\citenamefont {Freund}\ and\ \citenamefont
  {Suresh}(2004)}]{Freund:2004:}%
  \BibitemOpen
  \bibfield  {author} {\bibinfo {author} {\bibfnamefont {L.~B.}\ \bibnamefont
  {Freund}}\ and\ \bibinfo {author} {\bibfnamefont {S.}~\bibnamefont
  {Suresh}},\ }\href@noop {} {{\selectlanguage {English}\emph {\bibinfo {title}
  {Thin {{Film Materials}}: {{Stress}}, {{Defect Formation}} and {{Surface
  Evolution}}}}}}\ (\bibinfo  {publisher} {{Cambridge University Press}},\
  \bibinfo {year} {2004})\BibitemShut {NoStop}%
\bibitem [{\citenamefont {Gola}\ and\ \citenamefont
  {Pastewka}(2019)}]{gola_scratching_2019}%
  \BibitemOpen
  \bibfield  {author} {\bibinfo {author} {\bibfnamefont {A.}~\bibnamefont
  {Gola}}\ and\ \bibinfo {author} {\bibfnamefont {L.}~\bibnamefont
  {Pastewka}},\ }\href {\doibase 10/gf22x2} {\bibfield  {journal} {\bibinfo
  {journal} {Lubricants}\ }\textbf {\bibinfo {volume} {7}},\ \bibinfo {pages}
  {44} (\bibinfo {year} {2019})}\BibitemShut {NoStop}%
\bibitem [{\citenamefont {Khruschov}(1974)}]{Khruschov:1974:69}%
  \BibitemOpen
  \bibfield  {author} {\bibinfo {author} {\bibfnamefont {M.~M.}\ \bibnamefont
  {Khruschov}},\ }\href {\doibase 10.1016/0043-1648(74)90102-1} {\bibfield
  {journal} {\bibinfo  {journal} {Wear}\ }\textbf {\bibinfo {volume} {28}},\
  \bibinfo {pages} {69} (\bibinfo {year} {1974})}\BibitemShut {NoStop}%
\bibitem [{\citenamefont {Rigney}\ and\ \citenamefont
  {Hirth}(1979)}]{Rigney:1979:345}%
  \BibitemOpen
  \bibfield  {author} {\bibinfo {author} {\bibfnamefont {D.~A.}\ \bibnamefont
  {Rigney}}\ and\ \bibinfo {author} {\bibfnamefont {J.~P.}\ \bibnamefont
  {Hirth}},\ }\href {\doibase 10.1016/0043-1648(79)90087-5} {\bibfield
  {journal} {\bibinfo  {journal} {Wear}\ }\textbf {\bibinfo {volume} {53}},\
  \bibinfo {pages} {345} (\bibinfo {year} {1979})}\BibitemShut {NoStop}%
\bibitem [{\citenamefont {Mishra}\ and\ \citenamefont
  {Szlufarska}(2012)}]{Mishra:2012:417}%
  \BibitemOpen
  \bibfield  {author} {\bibinfo {author} {\bibfnamefont {M.}~\bibnamefont
  {Mishra}}\ and\ \bibinfo {author} {\bibfnamefont {I.}~\bibnamefont
  {Szlufarska}},\ }\href {\doibase 10.1007/s11249-011-9899-y} {\bibfield
  {journal} {\bibinfo  {journal} {Tribol. Lett.}\ }\textbf {\bibinfo {volume}
  {45}},\ \bibinfo {pages} {417} (\bibinfo {year} {2012})}\BibitemShut
  {NoStop}%
\bibitem [{\citenamefont {Mishra}\ \emph {et~al.}(2012)\citenamefont {Mishra},
  \citenamefont {Egberts}, \citenamefont {Bennewitz},\ and\ \citenamefont
  {Szlufarska}}]{Mishra:2012:45452}%
  \BibitemOpen
  \bibfield  {author} {\bibinfo {author} {\bibfnamefont {M.}~\bibnamefont
  {Mishra}}, \bibinfo {author} {\bibfnamefont {P.}~\bibnamefont {Egberts}},
  \bibinfo {author} {\bibfnamefont {R.}~\bibnamefont {Bennewitz}}, \ and\
  \bibinfo {author} {\bibfnamefont {I.}~\bibnamefont {Szlufarska}},\ }\href
  {\doibase 10.1103/PhysRevB.86.045452} {\bibfield  {journal} {\bibinfo
  {journal} {Phys. Rev. B}\ }\textbf {\bibinfo {volume} {86}},\ \bibinfo
  {pages} {045452} (\bibinfo {year} {2012})}\BibitemShut {NoStop}%
\bibitem [{\citenamefont {Luo}\ \emph {et~al.}(2015)\citenamefont {Luo},
  \citenamefont {Zhang},\ and\ \citenamefont {Schwaiger}}]{Luo:2015:67}%
  \BibitemOpen
  \bibfield  {author} {\bibinfo {author} {\bibfnamefont {Z.-P.}\ \bibnamefont
  {Luo}}, \bibinfo {author} {\bibfnamefont {G.-P.}\ \bibnamefont {Zhang}}, \
  and\ \bibinfo {author} {\bibfnamefont {R.}~\bibnamefont {Schwaiger}},\ }\href
  {\doibase 10.1016/j.scriptamat.2015.05.022} {\bibfield  {journal} {\bibinfo
  {journal} {Scr. Mater.}\ }\textbf {\bibinfo {volume} {107}},\ \bibinfo
  {pages} {67} (\bibinfo {year} {2015})}\BibitemShut {NoStop}%
\bibitem [{\citenamefont {Kim}\ \emph {et~al.}(2009)\citenamefont {Kim},
  \citenamefont {Karthikeyan},\ and\ \citenamefont {Rigney}}]{Kim:2009:1130}%
  \BibitemOpen
  \bibfield  {author} {\bibinfo {author} {\bibfnamefont {H.~J.}\ \bibnamefont
  {Kim}}, \bibinfo {author} {\bibfnamefont {S.}~\bibnamefont {Karthikeyan}}, \
  and\ \bibinfo {author} {\bibfnamefont {D.}~\bibnamefont {Rigney}},\ }\href
  {\doibase 10.1016/j.wear.2009.01.030} {\bibfield  {journal} {\bibinfo
  {journal} {Wear}\ }\textbf {\bibinfo {volume} {267}},\ \bibinfo {pages}
  {1130} (\bibinfo {year} {2009})}\BibitemShut {NoStop}%
\bibitem [{\citenamefont {Rigney}\ and\ \citenamefont
  {Karthikeyan}(2010)}]{Rigney:2010:3}%
  \BibitemOpen
  \bibfield  {author} {\bibinfo {author} {\bibfnamefont {D.~A.}\ \bibnamefont
  {Rigney}}\ and\ \bibinfo {author} {\bibfnamefont {S.}~\bibnamefont
  {Karthikeyan}},\ }\href {\doibase 10.1007/s11249-009-9498-3} {\bibfield
  {journal} {\bibinfo  {journal} {Tribol Lett}\ }\textbf {\bibinfo {volume}
  {39}},\ \bibinfo {pages} {3} (\bibinfo {year} {2010})}\BibitemShut {NoStop}%
\bibitem [{\citenamefont {Fu}\ \emph {et~al.}(2001)\citenamefont {Fu},
  \citenamefont {Falk},\ and\ \citenamefont {Rigney}}]{fu_sliding_2001}%
  \BibitemOpen
  \bibfield  {author} {\bibinfo {author} {\bibfnamefont {X.-Y.}\ \bibnamefont
  {Fu}}, \bibinfo {author} {\bibfnamefont {M.~L.}\ \bibnamefont {Falk}}, \ and\
  \bibinfo {author} {\bibfnamefont {D.~A.}\ \bibnamefont {Rigney}},\ }\href
  {\doibase 10.1016/S0043-1648(01)00607-X} {\bibfield  {journal} {\bibinfo
  {journal} {Wear}\ }\textbf {\bibinfo {volume} {250}},\ \bibinfo {pages} {420}
  (\bibinfo {year} {2001})}\BibitemShut {NoStop}%
\bibitem [{\citenamefont {Fu}\ \emph {et~al.}(2003)\citenamefont {Fu},
  \citenamefont {Rigney},\ and\ \citenamefont {Falk}}]{fu_sliding_2003}%
  \BibitemOpen
  \bibfield  {author} {\bibinfo {author} {\bibfnamefont {X.-Y.}\ \bibnamefont
  {Fu}}, \bibinfo {author} {\bibfnamefont {D.~A.}\ \bibnamefont {Rigney}}, \
  and\ \bibinfo {author} {\bibfnamefont {M.~L.}\ \bibnamefont {Falk}},\ }\href
  {\doibase 10.1016/S0022-3093(02)01999-3} {\bibfield  {journal} {\bibinfo
  {journal} {J. Non-Cryst. Solids}\ }\textbf {\bibinfo {volume} {317}},\
  \bibinfo {pages} {206} (\bibinfo {year} {2003})}\BibitemShut {NoStop}%
\bibitem [{\citenamefont {Wu}\ \emph {et~al.}(2005)\citenamefont {Wu},
  \citenamefont {Karthikeyan}, \citenamefont {Falk},\ and\ \citenamefont
  {Rigney}}]{wu_tribological_2005}%
  \BibitemOpen
  \bibfield  {author} {\bibinfo {author} {\bibfnamefont {J.~H.}\ \bibnamefont
  {Wu}}, \bibinfo {author} {\bibfnamefont {S.}~\bibnamefont {Karthikeyan}},
  \bibinfo {author} {\bibfnamefont {M.~L.}\ \bibnamefont {Falk}}, \ and\
  \bibinfo {author} {\bibfnamefont {D.~A.}\ \bibnamefont {Rigney}},\ }\href
  {\doibase 10.1016/j.wear.2004.11.028} {\bibfield  {journal} {\bibinfo
  {journal} {Wear}\ }\textbf {\bibinfo {volume} {259}},\ \bibinfo {pages} {744}
  (\bibinfo {year} {2005})}\BibitemShut {NoStop}%
\bibitem [{\citenamefont {Odunuga}\ \emph {et~al.}(2005)\citenamefont
  {Odunuga}, \citenamefont {Li}, \citenamefont {Krasnochtchekov}, \citenamefont
  {Bellon},\ and\ \citenamefont {Averback}}]{odunuga_forced_2005}%
  \BibitemOpen
  \bibfield  {author} {\bibinfo {author} {\bibfnamefont {S.}~\bibnamefont
  {Odunuga}}, \bibinfo {author} {\bibfnamefont {Y.}~\bibnamefont {Li}},
  \bibinfo {author} {\bibfnamefont {P.}~\bibnamefont {Krasnochtchekov}},
  \bibinfo {author} {\bibfnamefont {P.}~\bibnamefont {Bellon}}, \ and\ \bibinfo
  {author} {\bibfnamefont {R.~S.}\ \bibnamefont {Averback}},\ }\href {\doibase
  10/dtbrgs} {\bibfield  {journal} {\bibinfo  {journal} {Phys. Rev. Lett.}\
  }\textbf {\bibinfo {volume} {95}},\ \bibinfo {pages} {045901} (\bibinfo {year}
  {2005})}\BibitemShut {NoStop}%
\bibitem [{\citenamefont {Ashkenazy}\ \emph {et~al.}(2012)\citenamefont
  {Ashkenazy}, \citenamefont {Vo}, \citenamefont {Schwen}, \citenamefont
  {Averback},\ and\ \citenamefont {Bellon}}]{ashkenazy_shear_2012}%
  \BibitemOpen
  \bibfield  {author} {\bibinfo {author} {\bibfnamefont {Y.}~\bibnamefont
  {Ashkenazy}}, \bibinfo {author} {\bibfnamefont {N.~Q.}\ \bibnamefont {Vo}},
  \bibinfo {author} {\bibfnamefont {D.}~\bibnamefont {Schwen}}, \bibinfo
  {author} {\bibfnamefont {R.~S.}\ \bibnamefont {Averback}}, \ and\ \bibinfo
  {author} {\bibfnamefont {P.}~\bibnamefont {Bellon}},\ }\href {\doibase
  10/fhwfhm} {\bibfield  {journal} {\bibinfo  {journal} {Acta Mater.}\ }\textbf
  {\bibinfo {volume} {60}},\ \bibinfo {pages} {984} (\bibinfo {year}
  {2012})}\BibitemShut {NoStop}%
\bibitem [{\citenamefont {Zhou}\ \emph {et~al.}(2014)\citenamefont {Zhou},
  \citenamefont {Averback},\ and\ \citenamefont
  {Bellon}}]{zhou_stability_2014}%
  \BibitemOpen
  \bibfield  {author} {\bibinfo {author} {\bibfnamefont {J.}~\bibnamefont
  {Zhou}}, \bibinfo {author} {\bibfnamefont {R.~S.}\ \bibnamefont {Averback}},
  \ and\ \bibinfo {author} {\bibfnamefont {P.}~\bibnamefont {Bellon}},\ }\href
  {\doibase 10/f572c6} {\bibfield  {journal} {\bibinfo  {journal} {Acta
  Mater.}\ }\textbf {\bibinfo {volume} {73}},\ \bibinfo {pages} {116} (\bibinfo
  {year} {2014})}\BibitemShut {NoStop}%
\bibitem [{\citenamefont {Li}\ \emph {et~al.}(2019)\citenamefont {Li},
  \citenamefont {Pastewka},\ and\ \citenamefont {Gumbsch}}]{li_glass_2019}%
  \BibitemOpen
  \bibfield  {author} {\bibinfo {author} {\bibfnamefont {S.}~\bibnamefont
  {Li}}, \bibinfo {author} {\bibfnamefont {L.}~\bibnamefont {Pastewka}}, \ and\
  \bibinfo {author} {\bibfnamefont {P.}~\bibnamefont {Gumbsch}},\ }\href
  {\doibase 10/gft7mb} {\bibfield  {journal} {\bibinfo  {journal} {Acta
  Mater.}\ }\textbf {\bibinfo {volume} {165}},\ \bibinfo {pages} {577}
  (\bibinfo {year} {2019})}\BibitemShut {NoStop}%
\bibitem [{\citenamefont {{Hermann von
  Helmholtz}}(1868)}]{hermann_von_helmholtz_uber_1868}%
  \BibitemOpen
  \bibfield  {author} {\bibinfo {author} {\bibnamefont {{Hermann von
  Helmholtz}}},\ }\href@noop {} {\bibfield  {journal} {\bibinfo  {journal}
  {Ber. Akad. Wiss. Berlin}\ ,\ \bibinfo {pages} {215}} (\bibinfo {year}
  {1868})}\BibitemShut {NoStop}%
\bibitem [{\citenamefont {Thompson}(1871)}]{thompson_hydrokinetic_1871}%
  \BibitemOpen
  \bibfield  {author} {\bibinfo {author} {\bibfnamefont {W.}~\bibnamefont
  {Thompson}},\ }\href {\doibase 10.1080/14786447108640585} {\bibfield
  {journal} {\bibinfo  {journal} {The London, Edinburgh, and Dublin
  Philosophical Magazine and Journal of Science}\ }\textbf {\bibinfo {volume}
  {42}},\ \bibinfo {pages} {362} (\bibinfo {year} {1871})}\BibitemShut
  {NoStop}%
\bibitem [{\citenamefont {Thorpe}(1971)}]{Thorpe:1971:299}%
  \BibitemOpen
  \bibfield  {author} {\bibinfo {author} {\bibfnamefont {S.~A.}\ \bibnamefont
  {Thorpe}},\ }\href {\doibase 10.1017/S0022112071000557} {\bibfield  {journal}
  {\bibinfo  {journal} {J. Fluid Mech.}\ }\textbf {\bibinfo {volume} {46}},\
  \bibinfo {pages} {299} (\bibinfo {year} {1971})}\BibitemShut {NoStop}%
\bibitem [{\citenamefont {Pouryazdan}\ \emph {et~al.}(2017)\citenamefont
  {Pouryazdan}, \citenamefont {Kaus}, \citenamefont {Rack}, \citenamefont
  {Ershov},\ and\ \citenamefont {Hahn}}]{Pouryazdan:2017:1611}%
  \BibitemOpen
  \bibfield  {author} {\bibinfo {author} {\bibfnamefont {M.}~\bibnamefont
  {Pouryazdan}}, \bibinfo {author} {\bibfnamefont {B.~J.~P.}\ \bibnamefont
  {Kaus}}, \bibinfo {author} {\bibfnamefont {A.}~\bibnamefont {Rack}}, \bibinfo
  {author} {\bibfnamefont {A.}~\bibnamefont {Ershov}}, \ and\ \bibinfo {author}
  {\bibfnamefont {H.}~\bibnamefont {Hahn}},\ }\href {\doibase
  10.1038/s41467-017-01879-5} {\bibfield  {journal} {\bibinfo  {journal} {Nat.
  Commun.}\ }\textbf {\bibinfo {volume} {8}},\ \bibinfo {pages} {1611}
  (\bibinfo {year} {2017})}\BibitemShut {NoStop}%
\bibitem [{\citenamefont {Emge}\ \emph {et~al.}(2007)\citenamefont {Emge},
  \citenamefont {Karthikeyan}, \citenamefont {Kim},\ and\ \citenamefont
  {Rigney}}]{emge_effect_2007}%
  \BibitemOpen
  \bibfield  {author} {\bibinfo {author} {\bibfnamefont {A.}~\bibnamefont
  {Emge}}, \bibinfo {author} {\bibfnamefont {S.}~\bibnamefont {Karthikeyan}},
  \bibinfo {author} {\bibfnamefont {H.~J.}\ \bibnamefont {Kim}}, \ and\
  \bibinfo {author} {\bibfnamefont {D.~A.}\ \bibnamefont {Rigney}},\ }\href
  {\doibase 10.1016/j.wear.2007.01.095} {\bibfield  {journal} {\bibinfo
  {journal} {Wear}\ }\textbf {\bibinfo {volume} {263}},\ \bibinfo {pages} {614}
  (\bibinfo {year} {2007})}\BibitemShut {NoStop}%
\bibitem [{\citenamefont {Stoyanov}\ \emph {et~al.}(2013)\citenamefont
  {Stoyanov}, \citenamefont {Romero}, \citenamefont {Järvi}, \citenamefont
  {Pastewka}, \citenamefont {Scherge}, \citenamefont {Stemmer}, \citenamefont
  {Fischer}, \citenamefont {Dienwiebel},\ and\ \citenamefont
  {Moseler}}]{stoyanov_experimental_2013}%
  \BibitemOpen
  \bibfield  {author} {\bibinfo {author} {\bibfnamefont {P.}~\bibnamefont
  {Stoyanov}}, \bibinfo {author} {\bibfnamefont {P.~A.}\ \bibnamefont
  {Romero}}, \bibinfo {author} {\bibfnamefont {T.~T.}\ \bibnamefont {Järvi}},
  \bibinfo {author} {\bibfnamefont {L.}~\bibnamefont {Pastewka}}, \bibinfo
  {author} {\bibfnamefont {M.}~\bibnamefont {Scherge}}, \bibinfo {author}
  {\bibfnamefont {P.}~\bibnamefont {Stemmer}}, \bibinfo {author} {\bibfnamefont
  {A.}~\bibnamefont {Fischer}}, \bibinfo {author} {\bibfnamefont
  {M.}~\bibnamefont {Dienwiebel}}, \ and\ \bibinfo {author} {\bibfnamefont
  {M.}~\bibnamefont {Moseler}},\ }\href {\doibase 10.1007/s11249-012-0085-7}
  {\bibfield  {journal} {\bibinfo  {journal} {Tribol. Lett.}\ }\textbf
  {\bibinfo {volume} {50}},\ \bibinfo {pages} {67} (\bibinfo {year}
  {2013})}\BibitemShut {NoStop}%
\bibitem [{\citenamefont {Minowa}\ and\ \citenamefont
  {Sumino}(1992)}]{minowa_stress-induced_1992}%
  \BibitemOpen
  \bibfield  {author} {\bibinfo {author} {\bibfnamefont {K.}~\bibnamefont
  {Minowa}}\ and\ \bibinfo {author} {\bibfnamefont {K.}~\bibnamefont
  {Sumino}},\ }\href {\doibase 10.1103/PhysRevLett.69.320} {\bibfield
  {journal} {\bibinfo  {journal} {Phys. Rev. Lett.}\ }\textbf {\bibinfo
  {volume} {69}},\ \bibinfo {pages} {320} (\bibinfo {year} {1992})}\BibitemShut
  {NoStop}%
\bibitem [{\citenamefont {Pastewka}\ \emph {et~al.}(2011)\citenamefont
  {Pastewka}, \citenamefont {Moser}, \citenamefont {Gumbsch},\ and\
  \citenamefont {Moseler}}]{pastewka_anisotropic_2011}%
  \BibitemOpen
  \bibfield  {author} {\bibinfo {author} {\bibfnamefont {L.}~\bibnamefont
  {Pastewka}}, \bibinfo {author} {\bibfnamefont {S.}~\bibnamefont {Moser}},
  \bibinfo {author} {\bibfnamefont {P.}~\bibnamefont {Gumbsch}}, \ and\
  \bibinfo {author} {\bibfnamefont {M.}~\bibnamefont {Moseler}},\ }\href
  {\doibase 10.1038/nmat2902} {\bibfield  {journal} {\bibinfo  {journal} {Nat
  Mater}\ }\textbf {\bibinfo {volume} {10}},\ \bibinfo {pages} {34} (\bibinfo
  {year} {2011})}\BibitemShut {NoStop}%
\bibitem [{\citenamefont {Moras}\ \emph {et~al.}(2018)\citenamefont {Moras},
  \citenamefont {Klemenz}, \citenamefont {Reichenbach}, \citenamefont {Gola},
  \citenamefont {Uetsuka}, \citenamefont {Moseler},\ and\ \citenamefont
  {Pastewka}}]{moras_shear_2018}%
  \BibitemOpen
  \bibfield  {author} {\bibinfo {author} {\bibfnamefont {G.}~\bibnamefont
  {Moras}}, \bibinfo {author} {\bibfnamefont {A.}~\bibnamefont {Klemenz}},
  \bibinfo {author} {\bibfnamefont {T.}~\bibnamefont {Reichenbach}}, \bibinfo
  {author} {\bibfnamefont {A.}~\bibnamefont {Gola}}, \bibinfo {author}
  {\bibfnamefont {H.}~\bibnamefont {Uetsuka}}, \bibinfo {author} {\bibfnamefont
  {M.}~\bibnamefont {Moseler}}, \ and\ \bibinfo {author} {\bibfnamefont
  {L.}~\bibnamefont {Pastewka}},\ }\href {\doibase 10/gd3q47} {\bibfield
  {journal} {\bibinfo  {journal} {Phys. Rev. Mater.}\ }\textbf {\bibinfo
  {volume} {2}},\ \bibinfo {pages} {083601} (\bibinfo {year}
  {2018})}\BibitemShut {NoStop}%
\bibitem [{\citenamefont {Gola}\ and\ \citenamefont
  {Pastewka}(2018)}]{Gola:845721}%
  \BibitemOpen
  \bibfield  {author} {\bibinfo {author} {\bibfnamefont {A.}~\bibnamefont
  {Gola}}\ and\ \bibinfo {author} {\bibfnamefont {L.}~\bibnamefont
  {Pastewka}},\ }in\ \href@noop {} {\emph {\bibinfo {booktitle} {{{NIC
  Symposium}} 2018}}},\ \bibinfo {series} {NIC Series}, Vol.~\bibinfo {volume}
  {49}\ (\bibinfo  {publisher} {{Forschungszentrum J\"ulich GmbH}},\ \bibinfo
  {address} {J\"ulich},\ \bibinfo {year} {2018})\ pp.\ \bibinfo {pages}
  {247--254}\BibitemShut {NoStop}%
\bibitem [{\citenamefont {Berendsen}\ \emph {et~al.}(1984)\citenamefont
  {Berendsen}, \citenamefont {Postma}, \citenamefont {{van Gunsteren}},
  \citenamefont {DiNola},\ and\ \citenamefont {Haak}}]{Berendsen:1984:3684}%
  \BibitemOpen
  \bibfield  {author} {\bibinfo {author} {\bibfnamefont {H.~J.~C.}\
  \bibnamefont {Berendsen}}, \bibinfo {author} {\bibfnamefont {J.~P.~M.}\
  \bibnamefont {Postma}}, \bibinfo {author} {\bibfnamefont {W.~F.}\
  \bibnamefont {{van Gunsteren}}}, \bibinfo {author} {\bibfnamefont
  {A.}~\bibnamefont {DiNola}}, \ and\ \bibinfo {author} {\bibfnamefont {J.~R.}\
  \bibnamefont {Haak}},\ }\href {\doibase 10.1063/1.448118} {\bibfield
  {journal} {\bibinfo  {journal} {J. Chem. Phys.}\ }\textbf {\bibinfo {volume}
  {81}},\ \bibinfo {pages} {3684} (\bibinfo {year} {1984})}\BibitemShut
  {NoStop}%
\bibitem [{\citenamefont {Andersen}(1980)}]{andersen_molecular_1980}%
  \BibitemOpen
  \bibfield  {author} {\bibinfo {author} {\bibfnamefont {H.~C.}\ \bibnamefont
  {Andersen}},\ }\href {\doibase 10.1063/1.439486} {\bibfield  {journal}
  {\bibinfo  {journal} {J. Chem. Phys.}\ }\textbf {\bibinfo {volume} {72}},\
  \bibinfo {pages} {2384} (\bibinfo {year} {1980})}\BibitemShut {NoStop}%
\bibitem [{\citenamefont {Parrinello}\ and\ \citenamefont
  {Rahman}(1981)}]{Parrinello:1981:7182}%
  \BibitemOpen
  \bibfield  {author} {\bibinfo {author} {\bibfnamefont {M.}~\bibnamefont
  {Parrinello}}\ and\ \bibinfo {author} {\bibfnamefont {A.}~\bibnamefont
  {Rahman}},\ }\href {\doibase 10.1063/1.328693} {\bibfield  {journal}
  {\bibinfo  {journal} {J. Appl. Phys.}\ }\textbf {\bibinfo {volume} {52}},\
  \bibinfo {pages} {7182} (\bibinfo {year} {1981})}\BibitemShut {NoStop}%
\bibitem [{\citenamefont {Larsen}\ \emph {et~al.}(2016)\citenamefont {Larsen},
  \citenamefont {Schmidt},\ and\ \citenamefont
  {Schi\o{}tz}}]{Larsen:2016:55007}%
  \BibitemOpen
  \bibfield  {author} {\bibinfo {author} {\bibfnamefont {P.~M.}\ \bibnamefont
  {Larsen}}, \bibinfo {author} {\bibfnamefont {S.}~\bibnamefont {Schmidt}}, \
  and\ \bibinfo {author} {\bibfnamefont {J.}~\bibnamefont {Schi\o{}tz}},\
  }\href {\doibase 10.1088/0965-0393/24/5/055007} {\bibfield  {journal}
  {\bibinfo  {journal} {Modelling Simul. Mater. Sci. Eng.}\ }\textbf {\bibinfo
  {volume} {24}},\ \bibinfo {pages} {055007} (\bibinfo {year}
  {2016})}\BibitemShut {NoStop}%
\bibitem [{\citenamefont {Stukowski}(2010)}]{Stukowski:2010:15012}%
  \BibitemOpen
  \bibfield  {author} {\bibinfo {author} {\bibfnamefont {A.}~\bibnamefont
  {Stukowski}},\ }\href {\doibase 10.1088/0965-0393/18/1/015012} {\bibfield
  {journal} {\bibinfo  {journal} {Modelling Simul. Mater. Sci. Eng.}\ }\textbf
  {\bibinfo {volume} {18}},\ \bibinfo {pages} {015012} (\bibinfo {year}
  {2010})}\BibitemShut {NoStop}%
\bibitem [{\citenamefont {Falk}\ and\ \citenamefont
  {Langer}(1998{\natexlab{a}})}]{Falk:1998:7192}%
  \BibitemOpen
  \bibfield  {author} {\bibinfo {author} {\bibfnamefont {M.~L.}\ \bibnamefont
  {Falk}}\ and\ \bibinfo {author} {\bibfnamefont {J.~S.}\ \bibnamefont
  {Langer}},\ }\href {\doibase 10.1103/PhysRevE.57.7192} {\bibfield  {journal}
  {\bibinfo  {journal} {Phys. Rev. E}\ }\textbf {\bibinfo {volume} {57}},\
  \bibinfo {pages} {7192} (\bibinfo {year} {1998}{\natexlab{a}})}\BibitemShut
  {NoStop}%
\bibitem [{\citenamefont {Shimizu}\ \emph {et~al.}(2007)\citenamefont
  {Shimizu}, \citenamefont {Ogata},\ and\ \citenamefont
  {Li}}]{Shimizu:2007:2923}%
  \BibitemOpen
  \bibfield  {author} {\bibinfo {author} {\bibfnamefont {F.}~\bibnamefont
  {Shimizu}}, \bibinfo {author} {\bibfnamefont {S.}~\bibnamefont {Ogata}}, \
  and\ \bibinfo {author} {\bibfnamefont {J.}~\bibnamefont {Li}},\ }\href
  {\doibase 10.2320/matertrans.MJ200769} {\bibfield  {journal} {\bibinfo
  {journal} {Mater. Trans.}\ }\textbf {\bibinfo {volume} {48}},\ \bibinfo
  {pages} {2923} (\bibinfo {year} {2007})}\BibitemShut {NoStop}%
\bibitem [{\citenamefont {Gola}\ \emph {et~al.}(2019)\citenamefont {Gola},
  \citenamefont {Zhang}, \citenamefont {Pastewka},\ and\ \citenamefont
  {Schwaiger}}]{Gola:2019:}%
  \BibitemOpen
  \bibfield  {author} {\bibinfo {author} {\bibfnamefont {A.}~\bibnamefont
  {Gola}}, \bibinfo {author} {\bibfnamefont {G.-P.}\ \bibnamefont {Zhang}},
  \bibinfo {author} {\bibfnamefont {L.}~\bibnamefont {Pastewka}}, \ and\
  \bibinfo {author} {\bibfnamefont {R.}~\bibnamefont {Schwaiger}},\ }\href@noop
  {} {\  (\bibinfo {year} {2019})},\ \Eprint {http://arxiv.org/abs/1904.01942}
  {arXiv:1904.01942} \BibitemShut {NoStop}%
\bibitem [{\citenamefont {Falk}\ and\ \citenamefont
  {Langer}(1998{\natexlab{b}})}]{falk_dynamics_1998}%
  \BibitemOpen
  \bibfield  {author} {\bibinfo {author} {\bibfnamefont {M.~L.}\ \bibnamefont
  {Falk}}\ and\ \bibinfo {author} {\bibfnamefont {J.~S.}\ \bibnamefont
  {Langer}},\ }\href {\doibase 10.1103/PhysRevE.57.7192} {\bibfield  {journal}
  {\bibinfo  {journal} {Phys. Rev.E}\ }\textbf {\bibinfo {volume} {57}},\
  \bibinfo {pages} {7192} (\bibinfo {year} {1998}{\natexlab{b}})}\BibitemShut
  {NoStop}%
\bibitem [{\citenamefont {Li}\ \emph {et~al.}(2017)\citenamefont {Li},
  \citenamefont {Kreuter}, \citenamefont {Luo}, \citenamefont {Schwaiger},\
  and\ \citenamefont {Zhang}}]{Li:2017:20}%
  \BibitemOpen
  \bibfield  {author} {\bibinfo {author} {\bibfnamefont {X.}~\bibnamefont
  {Li}}, \bibinfo {author} {\bibfnamefont {T.}~\bibnamefont {Kreuter}},
  \bibinfo {author} {\bibfnamefont {X.~M.}\ \bibnamefont {Luo}}, \bibinfo
  {author} {\bibfnamefont {R.}~\bibnamefont {Schwaiger}}, \ and\ \bibinfo
  {author} {\bibfnamefont {G.~P.}\ \bibnamefont {Zhang}},\ }\href {\doibase
  10.1080/21663831.2016.1226976} {\bibfield  {journal} {\bibinfo  {journal}
  {Mater. Res. Lett.}\ }\textbf {\bibinfo {volume} {5}},\ \bibinfo {pages} {20}
  (\bibinfo {year} {2017})}\BibitemShut {NoStop}%
\bibitem [{\citenamefont {Li}\ \emph {et~al.}(2010)\citenamefont {Li},
  \citenamefont {Zhu}, \citenamefont {Zhang}, \citenamefont {Tan},
  \citenamefont {Wang},\ and\ \citenamefont {Wu}}]{Li:2010:3049}%
  \BibitemOpen
  \bibfield  {author} {\bibinfo {author} {\bibfnamefont {Y.~P.}\ \bibnamefont
  {Li}}, \bibinfo {author} {\bibfnamefont {X.~F.}\ \bibnamefont {Zhu}},
  \bibinfo {author} {\bibfnamefont {G.~P.}\ \bibnamefont {Zhang}}, \bibinfo
  {author} {\bibfnamefont {J.}~\bibnamefont {Tan}}, \bibinfo {author}
  {\bibfnamefont {W.}~\bibnamefont {Wang}}, \ and\ \bibinfo {author}
  {\bibfnamefont {B.}~\bibnamefont {Wu}},\ }\href {\doibase
  10.1080/14786431003776802} {\bibfield  {journal} {\bibinfo  {journal}
  {Philos. Mag.}\ }\textbf {\bibinfo {volume} {90}},\ \bibinfo {pages} {3049}
  (\bibinfo {year} {2010})}\BibitemShut {NoStop}%
\bibitem [{\citenamefont {Yan}\ \emph {et~al.}(2013)\citenamefont {Yan},
  \citenamefont {Zhu}, \citenamefont {Zhang},\ and\ \citenamefont
  {Yan}}]{Yan:2013:227}%
  \BibitemOpen
  \bibfield  {author} {\bibinfo {author} {\bibfnamefont {J.~W.}\ \bibnamefont
  {Yan}}, \bibinfo {author} {\bibfnamefont {X.~F.}\ \bibnamefont {Zhu}},
  \bibinfo {author} {\bibfnamefont {G.~P.}\ \bibnamefont {Zhang}}, \ and\
  \bibinfo {author} {\bibfnamefont {C.}~\bibnamefont {Yan}},\ }\href {\doibase
  10.1016/j.tsf.2012.11.052} {\bibfield  {journal} {\bibinfo  {journal} {Thin
  Solid Films}\ }\textbf {\bibinfo {volume} {527}},\ \bibinfo {pages} {227}
  (\bibinfo {year} {2013})}\BibitemShut {NoStop}%
\bibitem [{\citenamefont {Zaiser}(2006)}]{zaiser_scale_2006}%
  \BibitemOpen
  \bibfield  {author} {\bibinfo {author} {\bibfnamefont {M.}~\bibnamefont
  {Zaiser}},\ }\href {\doibase 10.1080/00018730600583514} {\bibfield  {journal}
  {\bibinfo  {journal} {Adv. Phys.}\ }\textbf {\bibinfo {volume} {55}},\
  \bibinfo {pages} {185} (\bibinfo {year} {2006})}\BibitemShut {NoStop}%
\bibitem [{\citenamefont {Hinkle}\ \emph {et~al.}(2019)\citenamefont {Hinkle},
  \citenamefont {Nöhring},\ and\ \citenamefont
  {Pastewka}}]{hinkle_universal_2019}%
  \BibitemOpen
  \bibfield  {author} {\bibinfo {author} {\bibfnamefont {A.~R.}\ \bibnamefont
  {Hinkle}}, \bibinfo {author} {\bibfnamefont {W.}~\bibnamefont {Nöhring}}, \
  and\ \bibinfo {author} {\bibfnamefont {L.}~\bibnamefont {Pastewka}},\
  }\href@noop {} {\  (\bibinfo {year} {2019})},\ \Eprint
  {http://arxiv.org/abs/1901.03236} {arXiv:1901.03236} \BibitemShut {NoStop}%
\bibitem [{\citenamefont {Bellon}\ and\ \citenamefont
  {Averback}(1995)}]{bellon_nonequilibrium_1995}%
  \BibitemOpen
  \bibfield  {author} {\bibinfo {author} {\bibfnamefont {P.}~\bibnamefont
  {Bellon}}\ and\ \bibinfo {author} {\bibfnamefont {R.~S.}\ \bibnamefont
  {Averback}},\ }\href {\doibase 10/bnsjhn} {\bibfield  {journal} {\bibinfo
  {journal} {Phys. Rev. Lett.}\ }\textbf {\bibinfo {volume} {74}},\ \bibinfo
  {pages} {1819} (\bibinfo {year} {1995})}\BibitemShut {NoStop}%
\bibitem [{\citenamefont {Gola}\ \emph {et~al.}(2018)\citenamefont {Gola},
  \citenamefont {Gumbsch},\ and\ \citenamefont {Pastewka}}]{GOLA2018236}%
  \BibitemOpen
  \bibfield  {author} {\bibinfo {author} {\bibfnamefont {A.}~\bibnamefont
  {Gola}}, \bibinfo {author} {\bibfnamefont {P.}~\bibnamefont {Gumbsch}}, \
  and\ \bibinfo {author} {\bibfnamefont {L.}~\bibnamefont {Pastewka}},\ }\href
  {\doibase https://doi.org/10.1016/j.actamat.2018.02.046} {\bibfield
  {journal} {\bibinfo  {journal} {Acta Mater.}\ }\textbf {\bibinfo {volume}
  {150}},\ \bibinfo {pages} {236} (\bibinfo {year} {2018})}\BibitemShut
  {NoStop}%
\bibitem [{\citenamefont {Zhang}\ \emph {et~al.}(2016)\citenamefont {Zhang},
  \citenamefont {Beyerlein}, \citenamefont {Zheng}, \citenamefont {Zhang},
  \citenamefont {Stukowski},\ and\ \citenamefont {Germann}}]{Zhang:2016:194}%
  \BibitemOpen
  \bibfield  {author} {\bibinfo {author} {\bibfnamefont {R.~F.}\ \bibnamefont
  {Zhang}}, \bibinfo {author} {\bibfnamefont {I.~J.}\ \bibnamefont
  {Beyerlein}}, \bibinfo {author} {\bibfnamefont {S.~J.}\ \bibnamefont
  {Zheng}}, \bibinfo {author} {\bibfnamefont {S.~H.}\ \bibnamefont {Zhang}},
  \bibinfo {author} {\bibfnamefont {A.}~\bibnamefont {Stukowski}}, \ and\
  \bibinfo {author} {\bibfnamefont {T.~C.}\ \bibnamefont {Germann}},\ }\href
  {\doibase 10.1016/j.actamat.2016.05.015} {\bibfield  {journal} {\bibinfo
  {journal} {Acta Mater.}\ }\textbf {\bibinfo {volume} {113}},\ \bibinfo
  {pages} {194} (\bibinfo {year} {2016})}\BibitemShut {NoStop}%
\bibitem [{\citenamefont {Anderson}\ \emph {et~al.}(2017)\citenamefont
  {Anderson}, \citenamefont {Hirth},\ and\ \citenamefont
  {Lothe}}]{hirth1982theory}%
  \BibitemOpen
  \bibfield  {author} {\bibinfo {author} {\bibfnamefont {P.~M.}\ \bibnamefont
  {Anderson}}, \bibinfo {author} {\bibfnamefont {J.~P.}\ \bibnamefont {Hirth}},
  \ and\ \bibinfo {author} {\bibfnamefont {J.}~\bibnamefont {Lothe}},\
  }\href@noop {} {\emph {\bibinfo {title} {Theory of Dislocations}}}\ (\bibinfo
   {publisher} {{John Wiley \& Sons}},\ \bibinfo {year} {2017})\BibitemShut
  {NoStop}%
\bibitem [{\citenamefont {Horstemeyer}\ \emph {et~al.}(2001)\citenamefont
  {Horstemeyer}, \citenamefont {Baskes},\ and\ \citenamefont
  {Plimpton}}]{horstemeyer_length_2001}%
  \BibitemOpen
  \bibfield  {author} {\bibinfo {author} {\bibfnamefont {M.~F.}\ \bibnamefont
  {Horstemeyer}}, \bibinfo {author} {\bibfnamefont {M.~I.}\ \bibnamefont
  {Baskes}}, \ and\ \bibinfo {author} {\bibfnamefont {S.~J.}\ \bibnamefont
  {Plimpton}},\ }\href {\doibase 10.1016/S1359-6454(01)00149-5} {\bibfield
  {journal} {\bibinfo  {journal} {Acta Mater.}\ }\textbf {\bibinfo {volume}
  {49}},\ \bibinfo {pages} {4363} (\bibinfo {year} {2001})}\BibitemShut
  {NoStop}%
\bibitem [{\citenamefont {Plimpton}(1995)}]{Plimpton:1995:1}%
  \BibitemOpen
  \bibfield  {author} {\bibinfo {author} {\bibfnamefont {S.}~\bibnamefont
  {Plimpton}},\ }\href {\doibase 10.1006/jcph.1995.1039} {\bibfield  {journal}
  {\bibinfo  {journal} {J. Comput. Phys.}\ }\textbf {\bibinfo {volume} {117}},\
  \bibinfo {pages} {1} (\bibinfo {year} {1995})}\BibitemShut {NoStop}%
\bibitem [{\citenamefont {Hjorth~Larsen}\ \emph {et~al.}(2017)\citenamefont
  {Hjorth~Larsen}, \citenamefont {Mortensen}, \citenamefont {Blomqvist},
  \citenamefont {Castelli}, \citenamefont {Christensen}, \citenamefont
  {Dułak}, \citenamefont {Friis}, \citenamefont {Groves}, \citenamefont
  {Hammer}, \citenamefont {Hargus}, \citenamefont {Hermes}, \citenamefont
  {Jennings}, \citenamefont {Bjerre~Jensen}, \citenamefont {Kermode},
  \citenamefont {Kitchin}, \citenamefont {Leonhard~Kolsbjerg}, \citenamefont
  {Kubal}, \citenamefont {Kaasbjerg}, \citenamefont {Lysgaard}, \citenamefont
  {Bergmann~Maronsson}, \citenamefont {Maxson}, \citenamefont {Olsen},
  \citenamefont {Pastewka}, \citenamefont {Peterson}, \citenamefont
  {Rostgaard}, \citenamefont {Schiøtz}, \citenamefont {Schütt}, \citenamefont
  {Strange}, \citenamefont {Thygesen}, \citenamefont {Vegge}, \citenamefont
  {Vilhelmsen}, \citenamefont {Walter}, \citenamefont {Zeng},\ and\
  \citenamefont {Jacobsen}}]{hjorth_larsen_atomic_2017}%
  \BibitemOpen
  \bibfield  {author} {\bibinfo {author} {\bibfnamefont {A.}~\bibnamefont
  {Hjorth~Larsen}}, \bibinfo {author} {\bibfnamefont {J.~J.}\ \bibnamefont
  {Mortensen}}, \bibinfo {author} {\bibfnamefont {J.}~\bibnamefont
  {Blomqvist}}, \bibinfo {author} {\bibfnamefont {I.~E.}\ \bibnamefont
  {Castelli}}, \bibinfo {author} {\bibfnamefont {R.}~\bibnamefont
  {Christensen}}, \bibinfo {author} {\bibfnamefont {M.}~\bibnamefont {Dułak}},
  \bibinfo {author} {\bibfnamefont {J.}~\bibnamefont {Friis}}, \bibinfo
  {author} {\bibfnamefont {M.~N.}\ \bibnamefont {Groves}}, \bibinfo {author}
  {\bibfnamefont {B.}~\bibnamefont {Hammer}}, \bibinfo {author} {\bibfnamefont
  {C.}~\bibnamefont {Hargus}}, \bibinfo {author} {\bibfnamefont {E.~D.}\
  \bibnamefont {Hermes}}, \bibinfo {author} {\bibfnamefont {P.~C.}\
  \bibnamefont {Jennings}}, \bibinfo {author} {\bibfnamefont {P.}~\bibnamefont
  {Bjerre~Jensen}}, \bibinfo {author} {\bibfnamefont {J.}~\bibnamefont
  {Kermode}}, \bibinfo {author} {\bibfnamefont {J.~R.}\ \bibnamefont
  {Kitchin}}, \bibinfo {author} {\bibfnamefont {E.}~\bibnamefont
  {Leonhard~Kolsbjerg}}, \bibinfo {author} {\bibfnamefont {J.}~\bibnamefont
  {Kubal}}, \bibinfo {author} {\bibfnamefont {K.}~\bibnamefont {Kaasbjerg}},
  \bibinfo {author} {\bibfnamefont {S.}~\bibnamefont {Lysgaard}}, \bibinfo
  {author} {\bibfnamefont {J.}~\bibnamefont {Bergmann~Maronsson}}, \bibinfo
  {author} {\bibfnamefont {T.}~\bibnamefont {Maxson}}, \bibinfo {author}
  {\bibfnamefont {T.}~\bibnamefont {Olsen}}, \bibinfo {author} {\bibfnamefont
  {L.}~\bibnamefont {Pastewka}}, \bibinfo {author} {\bibfnamefont
  {A.}~\bibnamefont {Peterson}}, \bibinfo {author} {\bibfnamefont
  {C.}~\bibnamefont {Rostgaard}}, \bibinfo {author} {\bibfnamefont
  {J.}~\bibnamefont {Schiøtz}}, \bibinfo {author} {\bibfnamefont
  {O.}~\bibnamefont {Schütt}}, \bibinfo {author} {\bibfnamefont
  {M.}~\bibnamefont {Strange}}, \bibinfo {author} {\bibfnamefont {K.~S.}\
  \bibnamefont {Thygesen}}, \bibinfo {author} {\bibfnamefont {T.}~\bibnamefont
  {Vegge}}, \bibinfo {author} {\bibfnamefont {L.}~\bibnamefont {Vilhelmsen}},
  \bibinfo {author} {\bibfnamefont {M.}~\bibnamefont {Walter}}, \bibinfo
  {author} {\bibfnamefont {Z.}~\bibnamefont {Zeng}}, \ and\ \bibinfo {author}
  {\bibfnamefont {K.~W.}\ \bibnamefont {Jacobsen}},\ }\href {\doibase
  10/f9wbtg} {\bibfield  {journal} {\bibinfo  {journal} {J. Phys. Condens.
  Matter}\ }\textbf {\bibinfo {volume} {29}},\ \bibinfo {pages} {273002}
  (\bibinfo {year} {2017})}\BibitemShut {NoStop}%
\end{thebibliography}


\end{document}